\documentstyle[11pt,paspconf,psfig]{article}

\begin{document}
\title{STELLAR VARIABILITY IN THE LOWER PART OF THE CLASSICAL INSTABILITY STRIP}
\author{E. Poretti, F. Musazzi}
\affil{Osservatorio di Brera, Via E. Bianchi 46, 22055 Merate, Italy}

\begin{abstract}
The photometric properties of the variable stars located in the lower part of
the classical instability strip are discussed. The importance of the 
determination of some light curve parameters and their connection with the
stellar models are stressed, with a particular emphasis on large amplitude
$\delta$ Sct stars.
\end{abstract}

%\keywords{}

\section{Introduction}
The lower part of the classical instability strip and the surrounding main
sequence are populated by a large variety of pulsating variable stars. In
the past years a great effort was made by the stellar
astronomers of Brera Observatory (E. Antonello, M. Bossi, 
L. Mantegazza, E. Poretti, F. Zerbi) to collect a large amount of
photometric and spectroscopic data on selected targets in order to 
improve stellar models. Observations were carried out
using the telescopes located in Merate (0.50 m and 1.02 m; a third telescope
having a diameter of 1.32 m will be newly available in 1998 after the mirror
installation) and at La Silla (European Southern Observatory, Chile, 
where we use the 0.50 m, 0.90 m, 1.0 m, and 1.4 m telescopes). In this 
paper we briefly
review the results obtained, with a particular emphasis on photometric ones,
since the current development of observational techniques renders them within
reach of well--equipped amateur astronomers.

\section{$\delta$ Sct stars}
The $\delta$ Sct class (DSCT in the GCVS notation) now contains most of the 
stars previously classified as $\delta$ Sct, SX Phe, AI Vel, and RR 
variables. The only discrimination now maintained is between Population I 
(i.e. DSCT stars) and
Population II (i.e. SXPHE stars). The amplitude of the light variation is
not considered a physical discriminant and hence both large (up to 0.50 mag) and
small (down to the mmag level) amplitude pulsators are included in the DSCT
class. For sake of clarity, let us consider separately the two subclasses.

\subsection{Large amplitude $\delta$ Sct stars}
Most of these variable stars are single--periodic, showing light curves 
significatively deviating from a pure sinusoid. In such a case it
is possible to fit the measurements by a sum of cosine functions having 
frequencies $f, 2f, 3f$, etc. (Fourier decomposition) and then to study the
particularities of the phase and amplitude parameters (see Pardo \& Poretti
1997 and references therein for the application of this technique to Cepheid
light curves). The particularities of the Fourier parameters of the $\delta$
Sct  light curves were reviewed by Poretti et al. (1990). Accurate
photoelectric photometry is available for all the single--periodic stars
 brighter than $V$ = 10. CCD photometry of faint objects is highly 
desirable in order
to increase the sample and to have a better definition of the properties of
the Fourier parameters.  In particular, it is important to confirm the 
bimodal
distribution of the amplitude ratio $R_{21}=A_{2f}/A_f$ and to verify if it is
possible to ascribe it to a resonance effect. Moreover, the phase parameter
$\phi_{21}=\phi_{2f}-\phi_f$ has been recognized as a powerful pulsation mode
discriminant and therefore to verify if all the large amplitude DSCT stars
are really fundamental radial mode pulsators, or if overtone pulsators can be
found among them. We recall here the case of V356 Aur (Poretti at al. 1987) and
AI Vel (Walraven et al. 1992), clearly indicating that multiperiodicity can be
observed and hence other modes than the fundamental can be excited.

In the figure two representative cases are presented. In the left half 
of the figure, the
light curves of V798 Cyg are shown.  The first period (upper panel) shows 
a rising branch covering more than half period, owing to a bump clearly visible
just after the minimum light. This kind of asymmetry is
quite uncommon in pulsating star light curves as discussed by Poretti \&
Antonello (1988). Regarding preliminary results described here, it 
should be noted that the new CCD measurements confirm not
only the existence of a  second term, but also that its frequency is
6.41~c~d$^{-1}$. This means that $f_1/f_2$=0.80, as happens for the other star
showing the same asymmetrical light curve, i.e. V1719 Cyg. The possible 
link between the double--mode pulsation and asymmetrical light curve deserves
further attention
in the near future. In the right half of the figure, the spectacular light 
curve of the 17th mag star V831 Tau
is presented.  The amplitude is about 0.70 mag and since the period is 
92 min, the ascending branch is only 20 min long!

\begin{figure*}
\psfig{figure=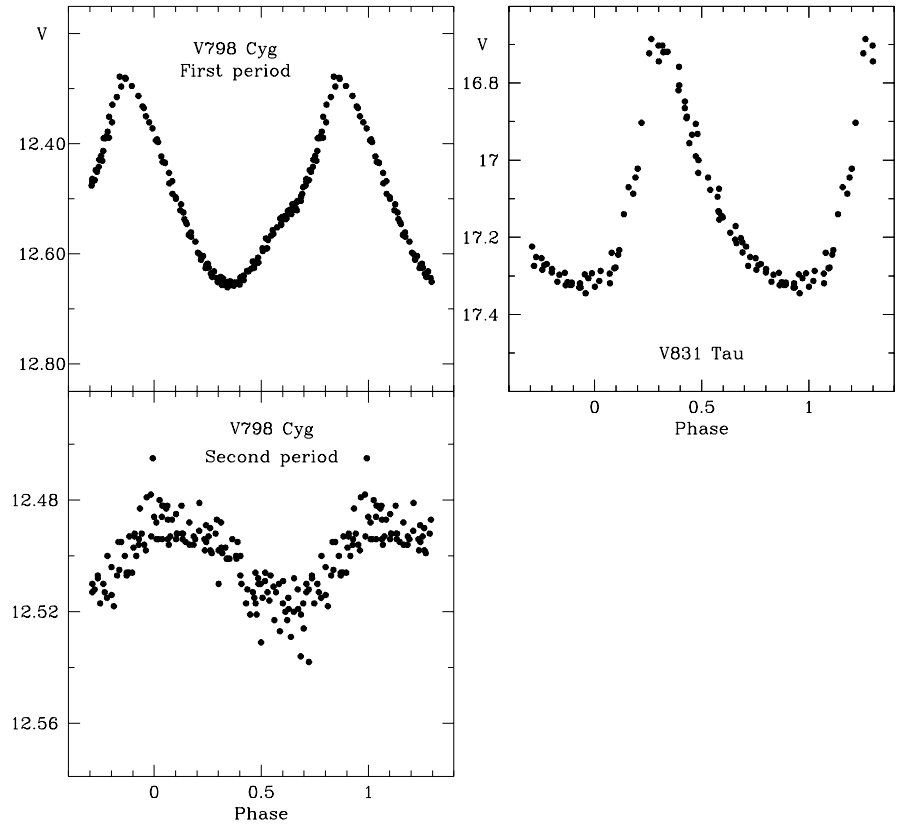,width=12.cm}
\caption{CCD photometry of large amplitude $\delta$ Sct stars. Left side:
the two periods observed  in the light curve of V798 Cyg.  Note the steeper
descending branch of the first period (upper panel) and the small amplitude
sinusoid of the second period (lower panel). Right side: the monoperiodic
star V831 Tau shows a very large amplitude and a short period (0.0643 d), 
making its light increase a spectacular phenomenon.}
\end{figure*}

\subsection{Small amplitude $\delta$ Sct stars}
In the small amplitude $\delta$ Sct stars we see the reverse case with 
respect to the large amplitude stars.  Multiperiodicity is quite common and 
monoperiodicity is rare. When the star is multiperiodic, it is
very complicated to make out the power spectrum owing to the interaction
between the excited modes and the spectral window. If measurements are performed
from a single site, not only the  excited frequency $f_1$ will be detected,
but also its aliases $f_1\pm n$ (where $n$ is an integral number of 
c~d$^{-1}$). When several
frequencies are simultaneously excited, it is not an easy task to correctly
separate the true frequencies from the aliases owing to the large number of
peaks visible in the power spectra. To simplify the analysis and to proceed
to an identification of a large number of excited modes $-$
{\it asteroseismology} $-$ it is necessary to
deal with a better spectral window, and this can be ensured by multisite
observations, i.e. carried out at different longitudes. In the case of a 
monoperiodic star, single--site measurements can provide a very useful check
of the constancy of the period (see Riboni et al. 1994 for the case of
$\beta$ Cas), but also the amplitude needs to be monitored (Poretti et al.
1996).
\section{$\gamma$ Dor stars}
In recent years a few early F--type stars showing small amplitude light 
variations
(a few hundredths of a mag) have been discovered, mainly because these stars
were used as comparison stars to measure $\delta$ Sct stars. Since
multiperiodicity
is currently observed (Balona et al. 1994; Poretti et al. 1997; Zerbi et al.
1997), $g$~--~mode nonradial pulsations are the most plausible cause of the
variation.  The existence of a such a class of pulsating variable
stars of spectral type $\sim$F0 is now accepted and $\gamma$ Dor has been
designated its prototype.
 The study of the light and line
profile variations will supply further support to the mode discrimination
by studying relationships between the phases of different curves (light,
colour, radial velocity, equivalent width, etc.). Powerful techniques have 
been recently developed to this end and applied by us to the study of
$p$--mode pulsators, i.e. the small amplitude $\delta$ Sct stars
(Mantegazza et al. 1994, 1996).

\section{Conclusions}
This rapid excursus among the photometric properties of the pulsators located
in the lower part of the instability strip discloses the possibilities offered
by the study of accurate light curves.
Since well--equipped amateur astronomers can achieve the required accuracy,
they can profitably contribute to the knowledge of stellar pulsation.
This possibility should not be disregarded by the AAVSO members, also considering
the tendency of the big observatories to close the small telescopes. In the
near future, the collaboration of amateur astronomers in the multisite 
campaigns will be more and more requested by professional researchers. 


\begin{references}
\reference Balona L., Krisciunas K., Cousins A.W.J., 1994, MNRAS, 270, 905
\reference Mantegazza L., Poretti E., Bossi M., 1994, \aap, 287, 95
\reference Mantegazza L., Poretti E., Bossi M., 1996, \aap, 312, 855
\reference Pardo I., Poretti, E. 1997, \aap, in press
\reference Poretti E., Antonello E., 1988, \aap, 199, 191
\reference Poretti E., Antonello E., LeBorgne J.F. 1990, \aap, 228, 350
\reference Poretti E., Koen C., Martinez P., Breuer P., de Alwis D., Haupt H.,
1997, MNRAS, in press
\reference Poretti E., Mantegazza L., Antonello E., 1987, \aap, 181, 273
\reference Poretti E., Mantegazza L., Bossi M., 1996, \aap, 312, 912
\reference Riboni E., Poretti E., Galli G., 1994, \aap~Suppl., 108, 55
\reference Walraven Th., Walraven J., Balona L., 1992, MNRAS, 254, 59
\reference Zerbi F., Garrido R., Rodriguez E., Krisciunas K., Crowe R.A.,
Roberts M., Guinan E.F., McCook G.P., Sperauskas J., Griffin R.F., 
Luedeke K.D., 1997, MNRAS, in press
\end{references}
\end{document}